\documentclass[sigconf]{acmart}
\AtBeginDocument{%
  }
\copyrightyear{2026}
\acmYear{2026}
\setcopyright{cc}
\setcctype{by}
\acmConference[MSR '26]{23rd International Conference on Mining Software Repositories}{April 13--14, 2026}{Rio de Janeiro, Brazil}
\acmBooktitle{23rd International Conference on Mining Software Repositories (MSR '26), April 13--14, 2026, Rio de Janeiro, Brazil}
\acmPrice{}
\acmDOI{10.1145/3793302.3793367}
\acmISBN{979-8-4007-2474-9/2026/04}

\usepackage{xspace}
\usepackage{cleveref}
\usepackage{listings}
\usepackage{subcaption}
\usepackage{soul}
\usepackage{mdframed} 

\usepackage[table]{xcolor}
\usepackage{booktabs}
\usepackage{siunitx}
\usepackage[most]{tcolorbox}
\usepackage{xcolor} 
\usepackage{soul} 
\usepackage{paralist}

\usepackage[normalem]{ulem}
\usepackage{textcomp}

\usepackage{float}

\usepackage{mdframed}

\usepackage{menukeys}

\newlength\MAX  
\newcommand*\Chart[1]{#1~\rlap{\textcolor{black!20}{\rule{\MAX}{2ex}}}\rule{#1\MAX}{2ex}}
\newcommand*\catChart[1]{\textbf{#1}~\rlap{\textcolor{black!20}{\rule{\MAX}{2ex}}}\rule{#1\MAX}{2ex}}

\newcommand{\greyrule}{\arrayrulecolor{black!30}\midrule\arrayrulecolor{black}}

\newcommand{\mysec}[1]{\vspace{0.08cm} \noindent \textbf{#1.}}

\usepackage{wrapfig}

\usepackage{ifthen}
\newboolean{showcomments}
\setboolean{showcomments}{true} 
\ifthenelse{\boolean{showcomments}}
  {\newcommand{\nb}[2]{
    \fcolorbox{gray}{yellow}{\bfseries\sffamily\scriptsize#1}
    {\sf\small$\blacktriangleright$\textit{#2}$\blacktriangleleft$}
  }
  
  }
  {\newcommand{\nb}[2]{}
  
  }

\newcommand{\ie}{i.e.,\xspace}

\definecolor{added}{HTML}{AAFFAA}
\definecolor{deleted}{HTML}{FFAAAA}
\definecolor{edited}{HTML}{FFDDB3}

\lstdefinestyle{normal}{
  language=python,
  basicstyle=\ttfamily\scriptsize,
  aboveskip=0pt,
  belowskip=0pt,
  escapeinside={(*}{*)},
}

\lstdefinestyle{a}{
  language=python,
  basicstyle=\ttfamily\scriptsize,
  aboveskip=0pt,
  belowskip=0pt,
  backgroundcolor=\color{added},
  escapeinside={(*}{*)},
}

\newtcbtheorem{Summary}{\bfseries Summary}{enhanced,drop shadow={black!50!white},
  coltitle=black,
  top=0.15in,
  attach boxed title to top left=
  {xshift=1.5em,yshift=-\tcboxedtitleheight/2},
  boxed title style={size=small,colback=white}
}{summary}

\newtcolorbox[auto counter]{prompt}[1][]{title={\bfseries},enhanced,drop shadow={black!50!white},
  coltitle=black,
  top=0.1in,
  attach boxed title to top left=
  {xshift=1.5em,yshift=-\tcboxedtitleheight/2},
  boxed title style={size=small,colback=white},}

\usepackage[normalem]{ulem}

\usepackage{tikz}

\definecolor{ABlue}{HTML}{127bca}
\definecolor{LHScolor}{HTML}{555555}

\usepackage[scaled]{helvet}
\usepackage[most]{tcolorbox}

\newcommand{\DOIbox}[1]{
\tcbsidebyside[
        bicolor,
        sidebyside,
        fontupper=\footnotesize\ttfamily\bfseries,
        fontlower=\footnotesize\sffamily\mdseries,
        nobeforeafter,
        shrink tight,
        extrude bottom by=1mm,
        sidebyside adapt=both,
        sidebyside gap=5pt,
        top=2pt,left=3pt,right=3pt,bottom=2pt,
        boxrule=0pt,
        rounded corners,
        coltext=white,
        colback=LHScolor,
        colbacklower=ABlue,
]{%
DOI
}{%
\href{#1}{#1}
}%
}

\begin{document}
\title{Beyond the Prompt: An Empirical Study of Cursor Rules}
\author{Shaokang Jiang}
\email{shj@uci.edu}
\affiliation{%
  \institution{University of California, Irvine}
  \city{Irvine}
  \state{California}
  \country{USA}
}

\author{Daye Nam}
\email{daye.nam@uci.edu}
\affiliation{%
  \institution{University of California, Irvine}
  \city{Irvine}
  \state{California}
  \country{USA}
}

\begin{abstract}
While Large Language Models (LLMs) have demonstrated remarkable capabilities, research shows that their effectiveness depends not only on explicit prompts but also on the broader context provided. 
This requirement is especially pronounced in software engineering, where the goals, architecture, and collaborative conventions of an existing project play critical roles in response quality. To support this, many AI coding assistants have introduced ways for developers to author persistent, machine-readable directives that encode a project's unique constraints. Although this practice is growing, the content of these directives remains unstudied.

This paper presents the first large-scale empirical study to characterize this emerging form of developer-provided context. Through a qualitative analysis of 401 open-source repositories with cursor rules, we developed a comprehensive taxonomy of project context that developers consider essential, organized into five high-level themes: Conventions, Guidelines, Project Information, LLM Directives, and Examples. Our study also explores how this context varies across different project types and programming languages, offering implications for the future context-aware AI developer tools.
\end{abstract}

\maketitle

\section{Introduction}

Research has consistently shown that the effectiveness of Large Language Models (LLMs) is highly
dependent not only on the direct prompts they receive~\cite{brown2020language} but also on the broader context in which
those prompts are situated~\cite{lewis2020retrieval, shaikh2024aligning}.
Context serves as the critical mechanism for guiding an LLM, influencing the quality and relevance
of its output and aligning the model with user expectations.

While this principle applies universally, the need for precise and structured context is particularly 
pronounced in software engineering~\cite{zhang2025coderag,wang2025coderag}. 
This distinction arises from the inherent nature of software development tasks. 
A developer's work is rarely self-contained; 
it is a goal-oriented task that must, in most cases, integrate into an existing project~\cite{begel2010codebook}. 
Furthermore, since most real-world software is built by collaborative teams, any new contribution 
must adhere to a shared set of conventions, including style guides, design patterns, and established 
processes, to ensure the codebase remains readable, consistent, and maintainable~\cite{bird2011sociotechnical, storey2014revolution}. 
Consequently, the context needed to guide an AI in these tasks could better be less about general knowledge 
but more about a rich set of prescriptive constraints around goal-oriented tasks.

\begin{figure}[t!]
\vspace{0.5cm}
\begin{mdframed}[backgroundcolor=gray!9,roundcorner=5pt]
    \small
\begin{verbatim}
This rule provides standards for frontend components:
  When working in components directory:
  - Always use Tailwind for styling
  - Use Framer Motion for animations
  - Follow component naming conventions

This rule enforces validation for API endpoints:
  In API directory:
  - Use zod for all validation
  - Define return types with zod schemas
  - Export types generated from schemas
\end{verbatim}
\end{mdframed}
\caption{Example cursor rule illustrating project conventions and usage constraints~\cite{cursor2025rules}}
\label{fig:cursor_rule_example}
\vspace{-1\baselineskip}
\end{figure}

Early research focused on helping developers prompt LLMs effectively for code generation~\cite{nam2025promptingllmscodeediting, yang2025prompts, pickering2025howhumans}. More recently, developer-authored, project-specific directives for AI agents have emerged as another approach~\cite{cursor2025rules,windsurf2025rules,copilot2025rules}. These ``rule files,'' exemplified by the \texttt{cursor.md} files for the Cursor IDE (see~\Cref{fig:cursor_rule_example}), are distinct from ephemeral, single-shot prompts; they are authored by end users for a machine collaborator to provide contextual guidance.

While the practice of authoring these AI directives is growing, their content remains unstudied.
Excessive or unoptimized context can lead to more complex and less accurate responses, as well as higher costs and latency. A lack of systematic understanding of what developers consider important, and what common patterns exist for their AI assistants, can result in ineffective use of these tools. Thus, developing such an understanding is crucial for improving the design and usability of AI coding assistants.

This paper presents the first large-scale empirical study to characterize this developer-provided context. We conducted a qualitative study of 401 open-source repositories containing developer-authored rule files to understand aspects developers consider essential when collaborating with AI. Then, we explored how this context varies across different project types and programming languages.

Our analysis reveals that developers provide a rich variety of context types, which we organized into five high-level themes: Convention, Guideline, Project, Example, and LLM Directive; and 20 detailed codes (\Cref{tab:cursor-rule-codes}). Many rules overlap with traditional software documentation practices, such as guidelines and project overviews, suggesting that developers are adapting established norms to AI-assisted development. Others, however, reflect patterns unique to AI directives, such as instructions targeting LLM behavior, which have no direct analog in human-oriented documentation and instead resemble prompt engineering techniques. The context provision also varies across different project types and programming languages, reflecting the diverse needs of developers when working with AI assistants. Our work can inform the design of future context-aware AI developer tools and help people better utilize AI coding assistants.

\section{Backgrounds and Related Work}

\subsection{Context Engineering}

In recent years, with the advancement of LLMs, context engineering has become a critical area of research. 
Many studies have shown that the performance of LLMs is highly dependent on how tasks are communicated, 
i.e., prompt engineering, and the context provided to the model~\cite{feldman2024ragged, kojima2022large, jiang2025questions}. 
Significant attention has been given to different prompting techniques, such as chain-of-thought prompting~\cite{wei2022chain, kojima2023large}, self-consistency~\cite{wang2023self}, zero-shot~\cite{kojima2022large}, and few-shot learning~\cite{brown2020language}. These techniques demonstrate that providing examples or instructions in the prompt can significantly improve model performance on various tasks. 
Retrieval-augmented generation~\cite{lewis2020retrieval, guu2020retrieval, shuster2021retrieval} also has been widely adopted to enhance context by incorporating relevant information from external sources, such as documents or databases, into the model's input. This approach allows LLMs to access a broader range of information and improve their performance on specific tasks. More recently, efforts have focused on creating more structured and persistent forms of context that can be easily utilized by the model during interactions. One promising direction is the development of memory-augmented models, which can maintain a long-term memory of past interactions and user preferences~\cite{shaikh2024aligning, zhang2018personalizing} to provide context-aware responses.

A similar idea has also been explored in the software engineering domain, where researchers have investigated various techniques to augment the context provided to LLMs for code generation and understanding tasks. 
For example, Mathews et al.~\cite{mathews2024test} found that including test cases enhances function-level code generation.
Jiang et al.~\cite{jiang2024self} showed that dividing the coding process into planning and implementation phases provides valuable guidance for LLMs. 
Similar ideas were demonstrated to be useful for code summarization~\cite{ahmed2024automatic}, code understanding~\cite{nam2024using}, repository-level code completion~\cite{li2024enhancing}, and program repair~\cite{autorecover}. 
RAG has also been applied in software engineering tasks, where relevant code snippets, documentation, and abstraction information are retrieved to supplement the model's context~\cite{zhong2025rag, zhang2025coderag, liu2024graphcoder, rag6}. This has allowed LLMs to access a broader range of information and improve their performance on code-related tasks. However, most approaches have focused on enhancing LLM performance in tasks that are well-defined and self-contained, such as code completion or code summarization, as covered by existing benchmarks.

\subsection{Rules for AI Coding Assistants}

Many AI coding assistants have introduced ways for developers to author system-level instructions that provide globally available context, preferences, or workflows for the AI to follow. These directives, often referred to as ``rule files,'' can include project-, team-, or user-specific guidelines that shape AI behavior. For example, Cursor, an AI-powered code editor, allows developers to create ``cursor rules'' in `.mdc' files specifying guidelines and constraints for code generation~\cite{cursor2025rules}. Because LLMs, by default, do not retain memory between sessions and may not know where to focus, these rules persistently encode important context, enabling AI assistants to incorporate them and generate more relevant and accurate suggestions.

Other AI coding assistants have also introduced similar features. For example, GitHub Copilot allows developers to define custom instructions that guide the behavior of the AI assistant. Several CLI tools, such as Claude CLI and Continue CLI, also integrate similar custom instruction features at the personal, project, and global levels. Some rules for these other AI coding assistants focus on integrating third-party tools, such as creating issues, running tests, or deploying code, into the agent workflow. In this paper, we focus solely on text-based rules that provide context for LLMs.

\begin{figure*}[!ht]
    \centering
    \includegraphics[width=0.9\linewidth]{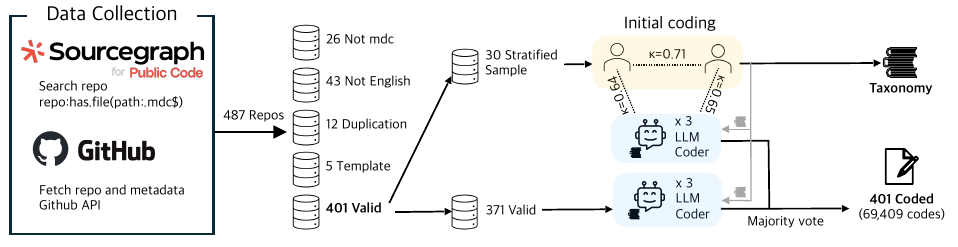} 
    \caption{Overview of the qualitative coding process for developing the taxonomy and coding rules for quantitative analysis.}
    \label{fig:coding-process}
\end{figure*}

\subsection{Documentation and Communication in Software Engineering}
Software engineering is an information-intensive and collaborative activity, making effective documentation and communication critical to the success of software projects. Numerous studies have investigated various aspects of software documentation, including the types of information developers seek~\cite{brandt2009two, maalej2013patterns, freund2015contextualizing, meng2018application, nam2024understanding}, as well as tools and techniques to improve documentation practices~\cite{robillard2011field, aghajani2019software, autocomplete2024jiang}. For example, Maalej and Robillard~\cite{maalej2013patterns} conducted a large-scale study of API documentation and identified common patterns of information needs among developers, finding twelve distinct types of information frequently sought in API documentation. Similarly, studies have explored the types of knowledge used in README files~\cite{prana2019categorizing, fronchetti2023contributing}, categorizing the information developers use when collaborating on open-source projects. These studies have provided valuable insights into the information needs of developers and the challenges they face in using documentation effectively. While we have a good understanding of the types of information developers need and provide when working with human collaborators, less is known about how they communicate with AI agents. Since AI coding agents have only recently become widely adopted and started to be considered collaborators, the ways developers document their needs for these agents needs further exploration. In this paper, we build upon prior work by focusing specifically on the context provided by developers for AI coding assistants, and how this context differs from traditional documentation practices.

\section{Methodology}

To understand the types of context developers provide when authoring cursor rules, we conducted a qualitative study of cursor rules from real-world open-source software repositories. An overview of our coding process is shown in \Cref{fig:coding-process}. We selected cursor rules because they were among the earliest to introduce this feature, are popular in the open-source community, and are frequently discussed in developer forums such as Reddit\footnote{\url{https://www.reddit.com/r/cursor/comments/1ju63ig/the_one_golden_cursor_rule_that_improved/}} and Hacker News\footnote{\url{https://news.ycombinator.com/item?id=43305919}}.

\subsection{Data Collection and Preprocessing}
\mysec{Data Fetching} 
We used Sourcegraph\footnote{\url{https://sourcegraph.com/}}, an open-source code search tool that indexes all public GitHub repositories with at least one star, representing the most popular open-source repositories on GitHub. We fetched all repositories containing a cursor rule file (\ie with the \texttt{.mdc} extension), excluding any forked or archived repositories, resulting in 487 repositories. We retrieved all repositories and metadata using GitHub's official API for further analysis.

\mysec{Data Cleaning} 
After cloning, we removed 15 repositories that contained no \texttt{.mdc} files. Manual inspection of the rest led to the removal of 11 whose \texttt{.mdc} files served unrelated purposes (e.g., SD card configuration) and 43 whose files were not written in English. We also removed 12 repositories that contained only empty data or exact duplicates. Finally, we excluded five repositories that provided only template or example \texttt{.mdc} files, as our study focuses on real-world usage, leaving 401 repositories for analysis.

\subsection{Dataset}

The 401 collected repositories containing cursor rules were created between 2008 and 2025, with the median creation date in 2023. The average repository size was 219.7 MB (SD = 727.6), with the largest repository being 12.63 GB. All collected repositories contained an average of 5,511.27 commits (SD = 11,201.06) per repository. Among all repositories, 291 (72.57\%) were created under an organization account, while the remaining 110 (27.43\%) were created under an individual account. The most commonly used licenses were MIT (152, 40.21\%), Apache-2.0 (76, 20.10\%), and NOASSERTION (76, 20.10\%). On average, each repository had 7,064 stars (SD = 21,509.10) and 916 forks (SD = 2,995.03). Across all repositories, the most commonly used programming languages were TypeScript (206, 51.37\%), Python (60, 14.96\%), and Go (24, 5.98\%).

Among all repositories, we identified a total of 1,876 \texttt{.mdc} files. On average, each repository contained 4.68 \texttt{.mdc} files (SD = 6.89), with a maximum of 56 \texttt{.mdc} files in a single repository. 
The latest cursor rule documentation recomments to place all rule files in the \texttt{.cursor} directory at the root of the repository. 
Among the 374 repositories (93.27\%) followed this recommendation, an average of 1.72 authors contributed to the directory (SD = 1.32), with a maximum of 8 authors. The average number of commits was 5.68 (SD = 9.31), with a maximum of 98 commits. Among the 281 repositories containing the cursor rule directory that has been modified after creation, people spent an average of 16 days (SD = 15.90) between each commit. The average length of all cursor rule files was 462.67 lines (SD = 1,197.00), with the longest file containing 11,076 lines.

\subsection{Qualitative Analysis}
\label{subsec:qualitative-analysis}
We qualitatively analyzed the content of cursor rules to understand the types of context developers provide when authoring them. We employed thematic analysis, following established procedures~\cite{braun2006using,clarke2013teaching}, to systematically identify recurring patterns and themes. The analysis was conducted iteratively, involving multiple rounds of coding, discussion, and refinement by the research team.

We focused our analysis on the textual content of cursor rules, excluding code blocks and other non-textual elements (e.g., images), as their purpose and content were generally self-explanatory. For instance, code blocks were typically included to provide illustrative code examples or snippets and therefore did not require further interpretation. Accordingly, we programmatically removed these non-textual elements to concentrate on the textual content.

We, the two authors, began by familiarizing ourselves with the data, reading through all cursor rules. Given that analyzing all rules from 401 repositories was impractical, we used stratified sampling to select 30 repositories from the entire pool for open coding. To ensure diversity, we sampled repositories across distinct topics assigned by each repository owner. We selected the top 10 most popular topics, which are "AI", "TypeScript", "tag-production", "Java", "security", "Prisma", "ecommerce", "MCP-Server", "PHP", and "DeepSeek". To prevent class imbalance, we randomly selected three repositories from each topic. We carefully checked the cursor rules, their descriptions, and related repository documentation to understand the context. During this initial review, we decided to code at the line level, as developers often expressed a single idea per line in cursor rule files. Since some lines were not directly related to context provision, such as section headers, we introduced "No Code" to indicate those lines. Through an inductive approach, we generated initial codes to describe the purpose and content of each cursor rule line. We recorded notes on emerging impressions and potential themes, referencing prior literature on documentation, developer communication, and context engineering to inform our perspective. Each researcher spent around 40 hours on this process.

After the in-depth review of the initial sample, we convened to discuss and refine the codes into a preliminary coding scheme. We then coded these repositories using the preliminary coding scheme again to ensure consistency and reliability. Discrepancies were discussed and resolved, and the coding scheme and definitions were further refined. We organized the codes into higher-level categories and themes through axial coding, iteratively improving the scheme through team consensus. Finally, independent coding of the initial 30 repositories by two researchers yielded substantial agreement~\cite{kappa, kappa1} (Cohen's Kappa = 0.71).

To scale the analysis to the full dataset of 401 repositories, we leveraged a large language model (LLM) to assist with the coding process, following prior work~\cite{LLMannotation1, cancode, cancode2, cancode3, cancode4}. We prompted the LLM with the final coding scheme and up to three representative examples for each code from our initial analysis, instructing it to assign codes to each line in the cursor rules with some surrounding context. We used \texttt{gemini-2.5-flash-preview-09-2025} with temperature 0.0 and dynamic thinking enabled. To enhance reliability, we repeated the labeling process three times and selected the majority vote as the final label for each line. If all three labels were different, we marked it as ``No Code'' to indicate uncertainty. 
To validate the LLM-assisted coding, we ran this process on the initial 30 repositories and found 
substantial agreement between the LLM and both human raters (Cohen's Kappa = 0.64 and 0.65), before applying it to the rest of the dataset.

\subsection{Threats to Validity}

Our research suffers from the usual threats of qualitative research in software engineering. While qualitative analysis enables a deep exploration and understanding of the data, it limits our ability to generalize beyond the studied dataset without further validation.

We focused exclusively on cursor rules, which may not represent other forms of developer-provided context for different coding agents. Additionally, our analysis was restricted to repositories that adopted cursor rules relatively early, potentially capturing developers' initial exploration of new project-level context provision, and may not reflect the sustained use of cursor rules, which could evolve as developers gain more experience with these tools.

A further limitation arises from using a large language model (LLM) to assist in coding. 
To improve reliability, we repeated the labeling process three times, selected the majority vote as 
the final label, and found substantial agreement with human coding. 
Nevertheless, the LLM may have introduced biases or misinterpretations, and failing to
capture the nuances, context, or intent behind developers' cursor rules. 
Such potential bias is a common limitation of qualitative analysis, especially when conducted at scale. 

Our dataset is limited to open-source repositories indexed by Sourcegraph, which focuses on popular GitHub projects. While this coverage is not exhaustive, it provides a reasonable sample of widely used repositories. Other rule formats, such as Copilot custom instructions, may include different types of contextual information. Since cursor rules were among the earliest to introduce this feature, have been widely adopted in the open-source community, and share similar underlying mechanisms with other rule formats, we believe our selection is sufficiently representative to provide initial insights into this emerging practice. Future work could study rules authored for other AI coding assistants to validate and extend our findings.

\section{Context Types in Cursor Rules}
\label{sec:context-types}

\begin{table*}[p]
    \centering
    \small
    \caption{Taxonomy of context types provided in cursor rules, including definitions and 
    representative examples. Highlighted codes are the high-level categories, that encompass
    the sub-codes listed below them.
    \% Repo indicates the proportion of repositories that included each 
    category and context type, and \% Code indicates the proportion of each category and context type
    across all cursor rule lines.}   
    \label{tab:cursor-rule-codes}
    \begin{tabular}{p{0.08\linewidth}p{0.7\linewidth} p{0.07\linewidth} p{0.07\linewidth}}
    \toprule
    \textbf{Code} & \textbf{Definition \& \textit{Example}} & \textbf{\% Repo}  & \textbf{\% Code} \\
    \midrule
    \colorbox[HTML]{F3AD5F}{\textbf{Project}} &  
    \begin{minipage}[t]{\linewidth}
        \textbf{Captures context that describes the software project of the repositories.}
    \end{minipage} &
    \catChart{0.85} & \catChart{0.35} \\ 

    \greyrule
    
    Env. &
    \begin{minipage}[t]{\linewidth}
        Specifies the technology stack (e.g., libraries, tools, hardware) and commands used to build, initialize, run, or test the project.\\
        \textit{Run the CLI `init` command: `npx trigger.dev@latest init`.}
    \end{minipage} &
    \Chart{0.72} & \Chart{0.07} \\ 

    \addlinespace[3pt]
    Func. &
    \begin{minipage}[t]{\linewidth}
        Describes the overall architecture, purpose, and key components of a software project, or the intended use of its various modules.\\
        \textit{`setup/`: Setting up the environment, installing dependencies.}
    \end{minipage} &
    \Chart{0.71} & \Chart{0.26} \\

    \addlinespace[3pt]
    Change &
    \begin{minipage}[t]{\linewidth}
        Describes shifts or updates in a project's technical implementation or configuration.\\
        \textit{No need for `postcss-import` or `autoprefixer` anymore}
    \end{minipage} &
    \Chart{0.17} & \Chart{0.01} \\

    \midrule
    \colorbox[HTML]{94C2FB}{\textbf{Convention}}  &  
    \begin{minipage}[t]{\linewidth}
        \textbf{Prescribed standards that define how code should be written and organized within a project.}
    \end{minipage} &
    \catChart{0.84} & \catChart{0.15} \\   
    \greyrule
    Code Style &
    \begin{minipage}[t]{\linewidth}
        Conventions for the code style, including naming, formatting, code writing, and other stylistic choices.\\
        \textit{Favor the use of functional components over class components.}
    \end{minipage} &
    \Chart{0.65} & \Chart{0.05} \\
   
    \addlinespace[3pt]
    \begin{minipage}[t]{\linewidth}Language/\\Framework \end{minipage} &
    \begin{minipage}[t]{\linewidth}
        Conventions specific to the project's language, framework, or library, such as preferred functions or constructs.\\
        \textit{Favor WordPress hooks (actions and filters) for extending functionality.}
    \end{minipage} &
    \Chart{0.64} & \Chart{0.06} \\

    \addlinespace[3pt]
    Structure &   \begin{minipage}[t]{\linewidth}
        Conventions for how files and directories are created, named, and organized within a software project.\\
        \textit{Use the `apps` directory for Next.js and Expo applications.}
    \end{minipage} &
   \Chart{0.59} & \Chart{0.04} \\

    \midrule
    \colorbox[HTML]{65A452}{\textbf{Guideline}}  &  
    \begin{minipage}[t]{\linewidth}
        \textbf{Prescriptive instructions to follow specific practices or avoid certain pitfalls.}
    \end{minipage} &
    \catChart{0.89} & \catChart{0.33} \\ 

    \greyrule
    QA &
    \begin{minipage}[t]{\linewidth}
        Practices for managing, preventing, and reporting errors, including logging, exception handling, and test requirements for verification.\\
        \textit{Comprehensive testing: Test all integration features.}
    \end{minipage} &
    \Chart{0.68} & \Chart{0.10} \\
    
    \addlinespace[3pt]
    General &
    \begin{minipage}[t]{\linewidth}
        High-level design and programming guidelines, including design patterns, separation of concerns, modularity, type safety, and code reuse.\\
        \textit{Maintain module separation of concerns.}
    \end{minipage} &
    \Chart{0.65} & \Chart{0.06} \\

    \addlinespace[3pt]
    Comm. &
    \begin{minipage}[t]{\linewidth}
        Guidelines for clear communication across the development workflow, including planning, implementation notes, PR documentation, and issue references.\\
        \textit{Include a clear title and description of the feature being implemented.}
    \end{minipage} &
    \Chart{0.64} & \Chart{0.07} \\
    \addlinespace[3pt]
    Perf. &
    \begin{minipage}[t]{\linewidth}
        Practices aimed at efficient resource usage, performance monitoring, and reuse with performance considerations.\\
        \textit{Cache expensive operations.}
    \end{minipage} &
    \Chart{0.43} & \Chart{0.03} \\
    \addlinespace[3pt]
    Consistency &
    \begin{minipage}[t]{\linewidth}
        Guidance to align new code with existing patterns, styles, structures, and logic.\\
        \textit{Follow existing patterns from similar components.}
    \end{minipage} &
    \Chart{0.38} & \Chart{0.01} \\
    \addlinespace[3pt]
        UI &
    \begin{minipage}[t]{\linewidth}
        Principles to improve usability, accessibility, and internationalization (i18n).\\
        \textit{All interactive elements (links, buttons, form controls, custom components) must be operable via a keyboard.}
    \end{minipage} &
    \Chart{0.34} & \Chart{0.02} \\
    \addlinespace[3pt]
    Security &
    \begin{minipage}[t]{\linewidth}
        Guidelines and practices to protect software from vulnerabilities and ensure data privacy.\\
        \textit{Implement proper nonce verification for form submissions.}
    \end{minipage} &
    \Chart{0.34} & \Chart{0.02} \\
    \addlinespace[3pt]
    Dependency &
    \begin{minipage}[t]{\linewidth}
        Practices to manage dependencies (e.g., avoid duplicates, manage versions) and to ensure compatibility.\\
        \textit{Ensure packages are properly isolated and dependencies are correctly managed.}
    \end{minipage} &
    \Chart{0.31} & \Chart{0.01} \\

    \midrule
    \colorbox[HTML]{AC77F8}{\textbf{LLM}}  &  
    \begin{minipage}[t]{\linewidth}
        \textbf{Directives that instruct LLMs on how to generate responses for the project.}
    \end{minipage} &
    \catChart{0.50} & \catChart{0.08} \\ 

    \greyrule
    
    Behavior &
    \begin{minipage}[t]{\linewidth}
        Directs LLMs to follow specific rules and guidelines when generating responses, e.g., to reference other documentation, use tools, or add verification steps.\\
        \textit{Always ask clarifying questions when you don't have full context or understanding of a task or question.}
    \end{minipage} &
    \Chart{0.36} & \Chart{0.02} \\ 
    
    \addlinespace[3pt]
    Workflow &
    \begin{minipage}[t]{\linewidth}
        Specifies a structured, multi-step process or orders of tasks an LLM must satisfy to complete a task.\\
        \textit{Update the scratchpad as progress is made (mark completed steps with [X])}
    \end{minipage} &
    \Chart{0.30} & \Chart{0.05} \\

    \addlinespace[3pt]
    Persona &
    \begin{minipage}[t]{\linewidth}
        Directives that instruct an LLM to adopt a specific role or persona while performing a task.\\
        \textit{You are an expert in Python, writing an AI application called chatgpt-mirai-qq-bot.}
    \end{minipage} &
    \Chart{0.18} & \Chart{0.00} \\

    \addlinespace[3pt]
    Formatting &
    \begin{minipage}[t]{\linewidth}
        Directs an LLM to generate output that strictly conforms to a specification such as a template, required format, language, or inclusion of references.\\
        \textit{Format should match: "Added/Improved/Fixed - Description of the change"}
    \end{minipage} &
    \Chart{0.16} & \Chart{0.01} \\

    \addlinespace[3pt]
    Granularity  &
    \begin{minipage}[t]{\linewidth}
        Directives instructing an LLM to adjust response detail.\\
        \textit{Avoid unnecessary verbosity and tangential remarks.}
    \end{minipage} &
    \Chart{0.12} & \Chart{0.00} \\

    \midrule
    
    \colorbox[HTML]{AAAAAA}{\textbf{Example}} &  
    \begin{minipage}[t]{\linewidth}
    \textbf{Demonstrates good and bad practices or provides a basic template.}
    \end{minipage} &
    \catChart{0.50} & \catChart{0.09} \\

    \addlinespace[3pt]
        \bottomrule
    \end{tabular}

\end{table*}

From our qualitative analysis, we identified five high-level context types commonly provided by developers: 
Project, Convention, Guideline, LLM Directive, and Example. 
Detailed definitions, representative examples of each context type, and their usage distributions are shown in~\Cref{tab:cursor-rule-codes}. In the following sections, we describe each category and its subcategories in detail, with representative examples.

\subsection{Project}
Most projects provided project-related context in their cursor rules. 
Codes in this category capture context that describes the software project, similar to the project overview and usage sections in traditional software documentation~\cite{maalej2013patterns,freund2015contextualizing}.

\mysec{Environment}
Many cursor rules describe the technology stack, including the programming languages, frameworks, libraries, and tools used. They also outline configuration steps, such as environment variables, configuration files, and required initialization commands. Most of this information overlaps with content typically found in the setup section of software documentation. The main purpose appears to be constraining the LLM to generate code compatible with the existing technology stack and helping LLM agents understand how to run tests, build the project, or deploy it.

\mysec{Functionality}
For functionality, this refers to descriptions of the overall architecture, purpose, and key components of a software project, or the intended use of its various modules. We observed many rules that explain different files or directories in the project, detailing their functionality and relationships to other components, often with links to those files or directories. This may be because developers want Cursor to properly reference those files. 

\mysec{Change}
We discovered that many cursor rules contain text describing recent updates to the project or its dependencies. Much of this content overlaps with what is typically found in release notes, such as \textit{``No need for `postcss-import` or `autoprefixer` anymore''}. These rules can help LLM agents avoid generating code that relies on deprecated features or conflicts with recent changes. This information is usually not obvious from the codebase itself.

\subsection{Convention}
The convention category includes rules that prescribe standards or practices for contributing to a project, typically based on a project-selected style guide or specific project usage requirement.

\mysec{Code Style}
Code style includes conventions related to code formatting, naming conventions, commenting practices, and other stylistic guidelines that developers should follow when writing code for the project. Many rules reference existing style guides, such as the Google Style Guide, Airbnb Style Guide, or PEP 8.

\mysec{Language/Framework Specific}
Language/framework usage includes conventions related to the specific programming languages, frameworks, or libraries used in the project, such as preferred idioms, patterns, or best practices. For example, \textit{``Use WordPress `@wordpress/element' instead of direct React import.''}

\mysec{Project Structure} 
Project structure includes conventions related to what developers need to follow when they want to make changes to the current project, such as directory structure, naming conventions for files and folders, and organization of modules or components. For example, \textit{``don't just create new directories and files that do not fit this pattern
otherwise Tanstack router''}. 

\subsection{Guideline}
The \textit{guideline} category captures context that provides high-level principles, best practices, and recommendations for developers to follow when contributing to a software project. Compared to conventions, guidelines are usually more general and abstract, covering common standards or best practices for areas not limited to a specific or a series of similar projects.

\mysec{Quality Assurance}
The most frequently mentioned code in the guideline was quality assurance, which includes instructions on testing, code reviews, debugging, and other practices to ensure the reliability and maintainability of the software. For example, \textit{``Utilize guard clauses to handle preconditions and invalid states early.''}

\mysec{General Programming}
Many cursor rules include very high-level programming best practices, such as separation of concerns, modularity, and type safety, which any experienced developer already knows. For example, \textit{``Maintain module separation of concerns''}.

\mysec{Communication}
Guidelines on documentation and communication were also among the most frequently mentioned codes in cursor rules. These rules provided recommendations on how to communicate effectively throughout the development workflow, such as writing clear commit messages, documenting code changes, and collaborating efficiently with team members.

\noindent\textbf{Performance} is used to provide guidance on efficient resource usage, such as ``Cache expensive operations.''
\textbf{Consistency} instructs agents to align with existing patterns, styles, and structures, such as ``Follow existing patterns...''
\textbf{UI} focuses on usability, accessibility, and internationalization, such as ``All interactive elements must be operable via a keyboard.''
\textbf{Security} addresses practices to mitigate vulnerabilities and protect data integrity, such as ``Implement proper nonce verification''.
And \textbf{dependency} provides guidance in package usage to ensure maintainability, such as ``Ensure packages are properly isolated and dependencies are correctly managed.''

\subsection{LLM Directive}
Another category we identified is \textit{LLM Directives}, which captures context that provides instructions and directives for LLMs to follow when responding. Unlike the previous three categories, which provide general information useful to any developer and have been widely discussed in prior literature~\cite{maalej2013patterns,freund2015contextualizing}, this category specifically targets LLM capabilities and offers tailored guidance.

Unlike the previous three categories, which are more prevalent in cursor rules, LLM Directives are less frequently mentioned. Only about 50\% of repositories (compared to 85\% for Project, 84\% for Convention, and 89\% for Guideline) contain LLM-specific instructions in their cursor rules, suggesting that many developers may not yet be familiar with the practice of authoring these prompt engineering strategies for LLMs~\cite{sahoo2024systematic,schulhoff2024prompt}.

\mysec{Behavior}
As the most frequently mentioned code in the \textit{LLM Directives} category, the behavior code provides instructions on how the LLM should behave when responding, such as being cautious, avoiding assumptions, verifying information, and asking clarifying questions. In more details, developers often uses a variety of strategies to guide LLM behavior in their projects. One common approach is structured reasoning, where prompts explicitly direct the model's thought process using techniques such as chain-of-thought prompting~\cite{wei2023chain, kojima2023large} or self-consistency~\cite{wang2023self}, helping the model arrive at more reliable outputs. Another frequent practice is asking for clarification: when prompts are ambiguous or underspecified, instructions may tell the LLM to request additional context, for instance, \textit{“Always ask clarifying questions when you don't have full context or understanding of a task or question.”} Many developers also include self-verification steps, requiring the LLM to check its assumptions, dependencies, or constraints before generating code, such as \textit{“Before generating any code, you MUST verify: 1. Are you importing from @trigger.dev/sdk/v3? If not, STOP and FIX.”} To prevent errors or undesired outcomes, instructions sometimes include specific prohibitions or antipatterns. Finally, developers frequently embed project-specific rules that are non-generalizable but essential for their particular context, for example, \textit{“Do not run test-teardown even if tests are successful—I will do that manually or request it explicitly.”} These practices collectively help ensure that LLMs behave predictably, safely, and in alignment with project requirements.

\mysec{Workflow}
The next most commonly mentioned theme was workflow, which specifies a structured, multistep process or sequence of tasks that an LLM should follow to complete a task, often by mimicking real-world human software development processes. Many of these workflows are designed to decompose complex tasks into smaller, manageable steps, such as \textit{``Before responding to any request, follow these steps: 1. Request Analysis, \ldots, 2. Solution Planning, \ldots''}. Some rules are written as conditional sequences, under which certain steps are executed only when specific conditions are met, for example, \textit{``When a user request begins with CHANGELOG:, add a changelog entry based on the provided prompt in the following files:''}.

\mysec{Persona}
Many cursor rules specify the persona the LLM should adopt, such as \textit{``You are an expert in Python, writing an AI application called chatgpt-mirai-qq-bot. It is a workflow-based chatbot system.''} Along with these persona definitions, some rules provide a high-level overview of the entire project, which the LLM is expected to be familiar with. We also observed that most persona instructions direct the LLM to assume technical roles, such as software engineer, with fewer projects emphasizing non-technical personas, such as story maker. Compared to behavior and workflow codes, persona instructions are less frequently mentioned in cursor rules.

\vspace{0.08cm}\noindent\textbf{Output Formatting} and \textbf{Granularity} are used to instruct the agent on how to respond in terms of format and level of detail. Formatting primarily directs the LLM on how to structure its responses, such as specifying templates, required formats, languages (e.g., \textit{``Respond in Chinese''}), or the inclusion of references. Granularity provides instructions on the level of detail the LLM should use in its responses, such as \textit{``Deliver the response in a minimal yet complete form. Avoid unnecessary verbosity and tangential remarks.''}

\subsection{Example}
We also identified one auxiliary content category, Example, which are typically provided alongside established rules, such as naming, formatting, or usage examples. For instance, \textit{``Example: \url{https://github.com/brainlid/langchain/pull/261}''}. \citet{maalej2013patterns} highlighted the importance of examples in helping developers. While we found a similar trend in cursor rules, the proportion of examples is relatively low in our quantitative results because code blocks were excluded (\Cref{subsec:qualitative-analysis}). Therefore, the number of example codes should be considered a lower bound.

\subsection{Discussion}
Overall, our qualitative analysis reveals that developers provide a diverse range of context types in cursor rules, spanning project-specific information, coding conventions, high-level guidelines, LLM-specific instructions, and examples. The prevalence of project-related context underscores developers' focus on conveying their software's unique aspects to LLMs. The widespread use of conventions and guidelines indicates developers' desire to maintain consistency and quality, aligning with established best practices. The presence of LLM-specific instructions, although less common, highlights the growing need to tailor LLM behavior to better suit the requirements of software development tasks.

We found that the spectrum of context types, aside from LLM-specific instructions, largely aligns with established software documentation practices~\cite{maalej2013patterns,freund2015contextualizing}. This suggests that developers are using cursor rules as an extension of traditional documentation, adapting familiar practices to the context of LLM-assisted coding. 

The relatively low adoption rate of LLM-specific instructions in cursor rules is noteworthy. One possible reason is that developers treat cursor rules as general background information for LLMs and may not recognize the importance of providing instructions tailored to the model. Another possibility is that developers are unsure whether such instructions are necessary or how to effectively communicate their needs and constraints to LLMs. Additionally, some may prefer LLMs to simply answer questions rather than generate code, providing only general context instead.

We also observed that developers in different domains may have different expectations and requirements when working with LLMs. For instance, frontend developers often focus on rapid iteration and want the LLM to automate more tasks. During our initial coding of cursor rules, we observed one frontend development repository (appsmith\footnote{\url{https://github.com/appsmithorg/appsmith}}) that provided logic for the LLM to automatically decide the workflow and complete development tasks step by step, with detailed linting and preflight checking code included in the cursor rules. At the time we collected the data and performed our analysis, cursor rules were not capable of executing such complex logic. However, as new features have been added to Cursor, such as hooks\footnote{\url{https://cursor.com/docs/agent/hooks}} released in late September, developers may now be able to define more advanced workflows with LLM assistants.

\section{Characteristics of Cursor Rules}

In \Cref{sec:context-types}, we identified different types of context that developers commonly provide in cursor rules. During this process, we observed that the context types may vary depending on various project characteristics, such as programming language and application domain. In this section, we further investigate these variations by exploring the following research questions, which are motivated by our observations during the coding process and by related literature on developer communication practices~\cite{maalej2013patterns,freund2015contextualizing}, project characteristics~\cite{plinfluencecontext}, and LLM prompt engineering strategies~\cite{sahoo2024systematic,schulhoff2024prompt}.

\vspace{\baselineskip}
\begin{mdframed}
\textbf{RQ1: How does context provision in cursor rules differ by programming language?}
\end{mdframed}
\vspace{0.5\baselineskip}
Prior research~\cite{plinfluencecontext, plinfluencecontext1} has shown that different programming languages influence developers' communication practices and the types of context they prioritize. For instance, statically typed languages such as Java, Rust, and TypeScript may require less explicit context regarding type safety, as their type systems enforce correctness at compile time. 
In contrast, dynamically typed languages such as Python and JavaScript could require more detailed context related to type checking and error handling, since such issues are only detectable at runtime. Furthermore, for recently adopted programming languages such as Rust~\cite{rustneeddifferent, rustneeddifferent1}, developers may need to provide additional context about language-specific features and idiomatic practices. 
In this question, we investigated whether these language-specific differences are actually reflected in those rules.

\vspace{\baselineskip}
\begin{mdframed}
\textbf{RQ2: How does context provision in cursor rules differ by application domain?}
\end{mdframed}
\vspace{0.5\baselineskip}

Previous work on best practices and conventions has shown that developers adhere to different sets of guidelines depending on the application domain of the software project~\cite{domainneedspecific1}. For example, in web development, best practices often emphasize responsive design, accessibility, and performance optimization, whereas embedded systems development prioritizes real-time constraints, resource management, and hardware interfacing~\cite{webreq,generaldomain}. 
With this research question, we investigated whether such domain-specific differences are also reflected in the context provided in cursor rules.

\begin{figure*}[t!]
    \centering
    \begin{minipage}{0.31\linewidth}
        \includegraphics[width=\linewidth]{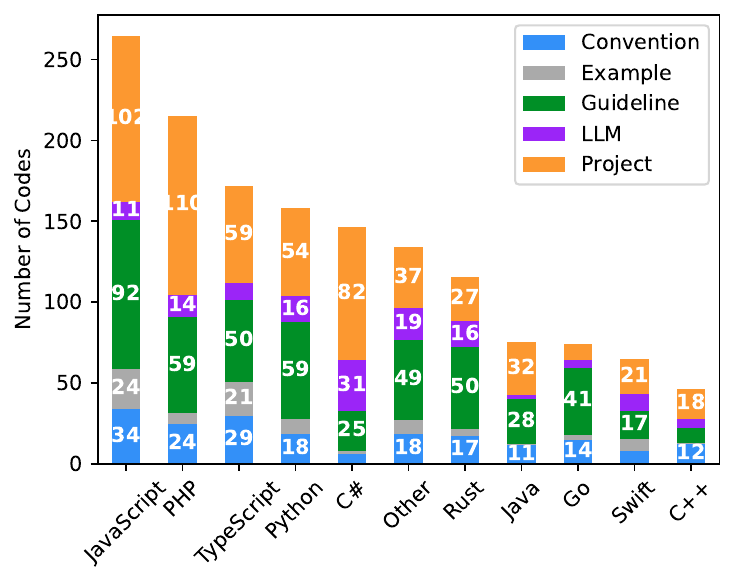}
        \caption{Distribution of the average number of codes for each context type by programming language.}
        \label{fig:context-type-language-distribution}
    \end{minipage}\hfill
    \begin{minipage}{0.31\linewidth}
         \includegraphics[width=\linewidth]{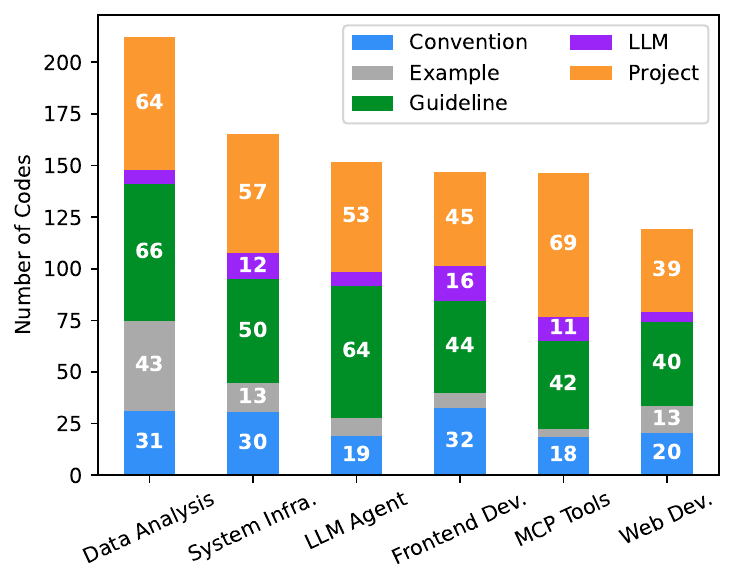}
        \caption{Distribution of the average number of codes for each context type by application domain.}
        \label{fig:context-type-topic-distribution}
    \end{minipage}\hfill
    \begin{minipage}{0.31\linewidth}
        \includegraphics[width=\linewidth]{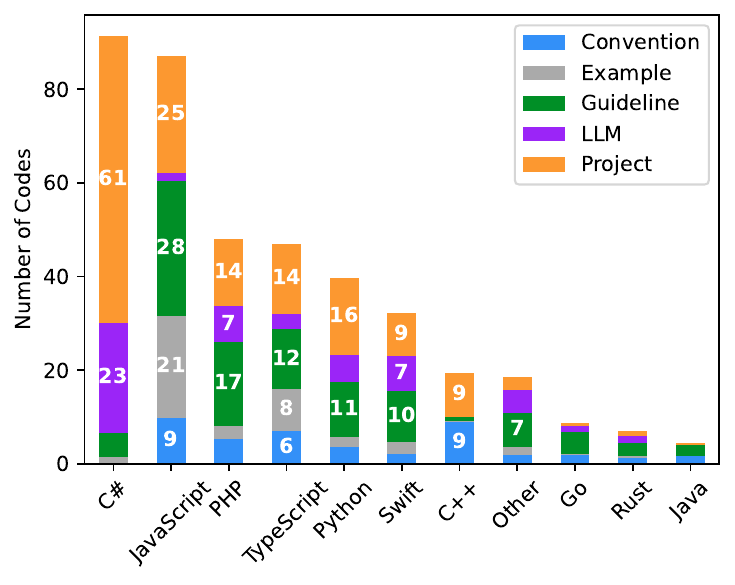}
        \caption{Distribution of the average number of duplicated lines for each context type by programming language.}
        \label{fig:duplication-language-distribution}
    \end{minipage}
\end{figure*}

\vspace{\baselineskip}
\begin{mdframed}
\textbf{RQ3: To what extent do developers reuse cursor rules?}
\end{mdframed}
\vspace{0.5\baselineskip}

In software engineering, reusability is a key principle that promotes efficiency and consistency~\cite{copypaste}. Developers are encouraged to reuse existing code snippets, libraries, and frameworks to expedite development and maintain consistency across projects. Similarly, prior work on prompt engineering has highlighted the benefits of reusing effective prompts from external sources or previous projects to enhance LLM performance~\cite{prompt11}. In our analysis, we observed that developers often reuse rules from existing documentation, coding standards, and best practices when authoring cursor rules. Additionally, we found several community-maintained repositories providing template cursor rules, which encourage the reuse of existing cursor rules. 
With this research question, we aimed to understand the extent to which developers reuse cursor rules from external sources or prior projects when authoring their own rules.

\vspace{\baselineskip}
\begin{mdframed}
\textbf{RQ4: To what extent do developers provide multiple types of context in cursor rules?}
\end{mdframed}
\vspace{0.5\baselineskip}
Many guidelines for traditional documentation recommend that developers provide multiple types of information to give a comprehensive overview of the project. For example, Maalej and Robillard~\cite{maalej2013patterns} suggest that effective documentation should include both high-level overviews and low-level implementation details to support various developer needs. Similarly, many studies on documentation have found that developers face challenges when documentation lacks code examples or explanations of the rationale behind design decisions~\cite{sohan2017study, ebert2021exploratory}. 
In RQ4, we investigated whether developers follow these best practices when authoring cursor rules.

\vspace{\baselineskip}
\begin{mdframed}
\textbf{RQ5: How does the timing of cursor rule addition relative to repository creation affect the context provided?} 
\end{mdframed} 
\vspace{0.5\baselineskip}

Since writing documentation for LLMs is a new practice, we hypothesized that the timing of cursor rule addition may influence the type of context included. In older repositories, developers might reuse existing project documentation rather than create detailed context for LLMs, whereas in newer repositories, they may provide more high-level, LLM-specific instructions to guide code generation. Therefore, we want to understand how the timing of rule addition relative to repository creation affects the context provided.

\subsection{RQ1: Programming Language}
\label{rq1}

To answer RQ1, we first identified the primary programming language used in each repository
with value provided by GitHub's API, and analyzing the distribution of context types across different programming languages.
We only included those programming languages that were top 10 in our dataset.
To make the result more reliable, we only included programming languages that had at least 6 repositories, resulting in the following top 10 languages: Typescript (206 repositories),  Python (60), Go (24), JavaScript (17),  Rust (16), PHP (10), Java (9), Swift (8), C\# (7), and C++ (6).

\mysec{Findings} Different programming languages exhibit different distributions of context types, as shown in \cref{fig:context-type-language-distribution}.

Developers tend to provide less context when working with statically typed languages such as Go, C\#, and Java, 
and provide more context when working with dynamic languages like JavaScript and PHP. 
This suggests developers expect stricter type checking in statically typed languages helps LLMs infer more from the code, reducing the need for extra context~\cite{typeisgood1,typeisgood}. 
For example, repositories of JavaScript as main languages tend to provide more guideline-related context (\textbf{M}ean=92) and more project-related context (M=102),
while those using TypeScript, a statically typed superset of JavaScript, 
has a lower average number of guideline-related context (M=50), and project-related context (M=59). 

Developers provided more overall context (M=207) for PHP projects compared to other languages,
possibly because PHP is an older language with a wide variety of coding styles and practices,
which may make it challenging for LLMs to infer the appropriate conventions and best practices without additional context.

Some less dominant, domain specific languages, such as C\#, tend to have more LLM-related context (M=31) in cursor rules. 
This may be because C\# is often used in enterprise applications and game development, where LLMs need to be more cautious about the generated code to avoid potential issues. Additionally, its similarity in structure to Java and other object-oriented languages—which are more commonly used and prevalent in LLM training data—may increase the necessity of providing additional context.

Among all languages, JavaScript and TypeScript most commonly include examples in cursor rules (M=24 and M=21). This likely reflects their use in frontend development with rapidly evolving frameworks and libraries, leading developers to provide examples to help LLMs capture specific usage patterns and best practices.

\mysec{Limitations}
Our analysis considered only the dominant programming language of each repository, as the GitHub API provides a single primary language. This may oversimplify the analysis of multi-language projects, such as full-stack web development projects. Additionally, our dataset is skewed, with over 50\% of repositories using TypeScript or JavaScript. Nevertheless, relative comparisons across languages should still hold, and this reflects the current adoption of cursor rules on GitHub. Further studies are necessary to better understand the differences and requirements when working with LLMs across programming languages.

\subsection{RQ2: Application Domain}

To identify application domains, we used the list of GitHub topics via the official GitHub API. 
Since each repository can have multiple topics and owners can assign them freely, we performed clustering over these topics using BERTopic~\cite{berttopic}. Before clustering, we removed generic terms such as ``ai'' and ``open source'', excluded non-English topics, and filtered out repositories labeled with only a single topic, resulting in 317 repositories for clustering.

To find the optimal clustering model, we conducted a greedy search over BERTopic hyperparameters to balance the number of topics, topic interpretability, the proportion of unclustered items, item-topic coherence (measured using the c\_v metric with Gensim~\cite{c-v1,c-v2}), and topic diversity (measured as the ratio of unique tokens to total document length). The final model produced six topics, achieving a low unclustered ratio (0.3\%), acceptable coherence (0.52), and high topic diversity (0.93). The research team further validated topic meaningfulness by manually inspecting top keywords and representative repositories. Given that c\_v is corpus-dependent~\cite{lim2024aligning} and developer-assigned topics are highly diverse, these results indicate a satisfactory clustering outcome.

Finally, we used BERTopic with UMAP~\cite{umap} (15 neighbors) for dimensionality reduction and HDBSCAN~\cite{hdbscan} for clustering, which identifies noise points without requiring a predefined number of clusters. We set the minimum cluster size to 20 (5\% of the dataset) and the minimum samples to 2 to ensure at least two repositories per topic. For embeddings, we used the officially recommended \texttt{all-MiniLM-L6-v2} model~\cite{all-MiniLM-L6-v2}. With excluding any unclustered repositories, we identified six distinct topics: 
\begin{inparadesc}
    \item \textbf{Web Development} (70 repositories), general-purpose web projects.
    \item \textbf{LLM Agents} (62 repositories), projects that build or evaluate LLM-based agents.
    \item \textbf{System Infrastructure} (61 repositories), full-stack development and system design, including domain-specific applications such as e-commerce (e.g., OpenCut).
    \item \textbf{Data Analysis} (50 repositories), projects focused on data analysis and data management tasks.
    \item \textbf{Frontend Development} (44 repositories), frontend tools, design systems, and user interface development (e.g., ant-design-web3).
    \item \textbf{MCP Tools} (29 repositories), plugins and extensions for coding assistants.
\end{inparadesc}

\mysec{Findings}
Figure~\ref{fig:context-type-topic-distribution} presents the distribution of the average number of codes
in each context category across detected domains. Overall, we found that different application domains tend to have different distributions of context types. 

More complex projects, such as data analysis, web development, and system infrastructure building, tend to provide more examples for LLMs (M=43, M=13, and M=13). These projects often involve intricate workflows, diverse data sources, complex component interactions, and extensive use of existing libraries, which may be difficult for LLMs to fully grasp without concrete examples.

For domains with few publicly available, up-to-date APIs (e.g., MCP code server projects), developers provide more project context to help LLMs understand project requirements (M = 69)~\cite{apihard}, whereas for domains with abundant online resources (e.g., web development), developers provide less project context (M = 39).

Developers tend to provide more guideline-related context when working on projects that involve unpredictable or difficult-to-test tasks requiring careful consideration before execution, such as LLM agent projects and data analysis projects (M=64 and M=66).

\mysec{Limitations}
It is possible that our topic clustering have misclassified some repositories, potentially affecting 
the analysis of domain topics and context types. 
We mitigated the risk of misclassification by manually reviewing the top keywords and representative repositories
for each topic to ensure the quality of our clustering.

Application domain and programming language are inherently strongly correlated. 
For example, frontend projects predominantly use JavaScript or TypeScript, whereas data analysis projects often rely on Python or R. 
This correlation may confound our analysis of application domains, as programming language also influences the types of context provided (\Cref{rq1}). 
However, such confounding is common in software engineering research, and disentangling the effects of programming language and application domain is inherently challenging. 
Despite this threat, we observed insights not solely attributable to programming language, 
such as the greater prevalence of guideline-related context in LLM agent and data analysis projects, 
showing the value of conducting a separate analysis on application domains.

\subsection{RQ3: Proportion of Duplicated Lines}

During our initial coding, we observed that a large portion of cursor rules appeared to be directly 
copied from other documents within similar repositories or from shared template repositories, 
such as ``awesome-cursor-rules''\footnote{\url{https://github.com/PatrickJS/awesome-cursorrules}}. 
To quantitatively assess the extent of such duplication, we analyzed the proportion of exactly 
duplicated lines across all repositories in our dataset.

\begin{figure*}[t!]
    \centering
    \begin{minipage}[b]{0.34\linewidth}
        \includegraphics[width=\linewidth]{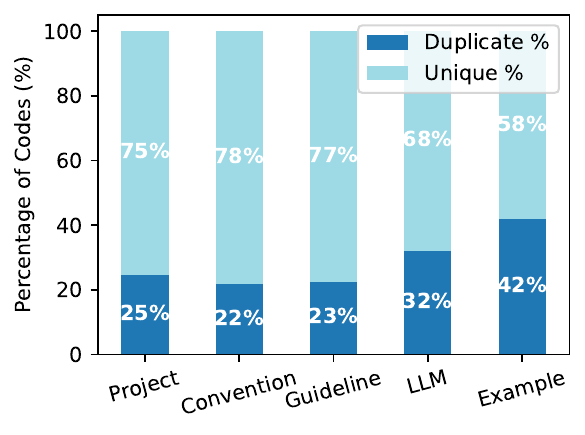}
        \caption{Percentage of code that are duplicated by context category.}
        \label{fig:bar-duplicated-lines}
    \end{minipage}\hfill
    \begin{minipage}[b]{0.205\linewidth}
        \includegraphics[width=\linewidth]{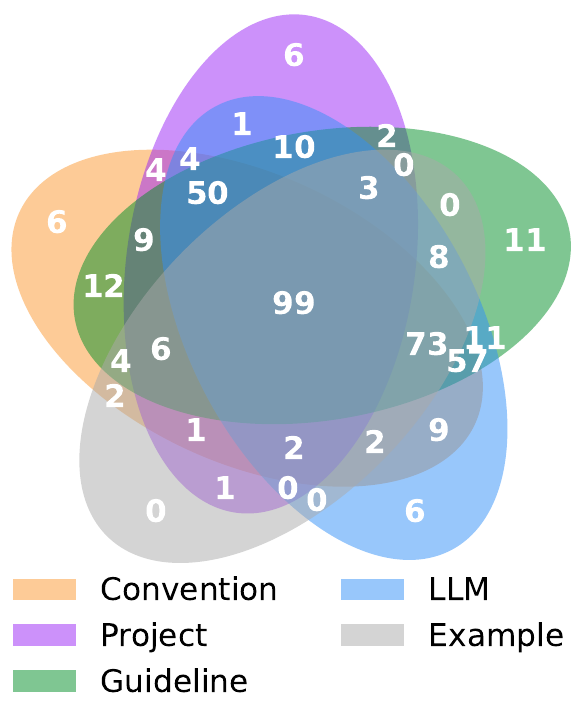}
        \caption{Venn of context type co-occurrences.}
        \label{fig:context-type-venn}
    \end{minipage}\hfill
    \begin{minipage}[b]{0.4\linewidth}
    \includegraphics[width=\linewidth]{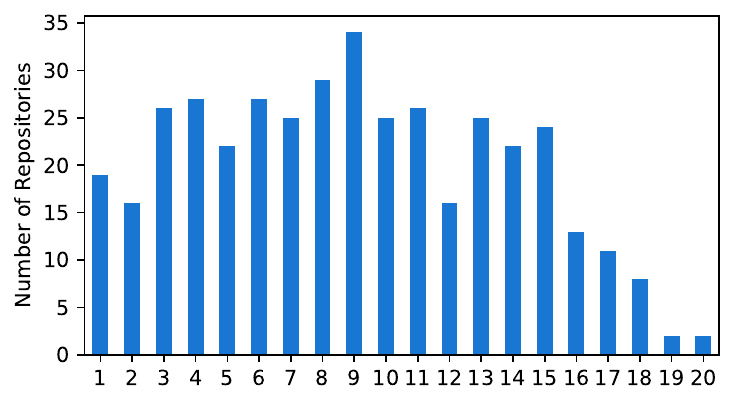}
    \caption{Distribution of the number of unique code types provided by repositories.}
    \label{fig:context-type-correlation}
    \end{minipage}
\end{figure*}

\mysec{Findings}
Overall, we identified a total of 19,917 (28.70\%) duplicated lines across all repositories with an average of 49.67 duplicate lines (SD = 164.20) per repository.
Figure~\ref{fig:bar-duplicated-lines} shows the percentage of code that are duplicated by context category.

Developers usually provide unique information about their own projects. We found that the proportion of duplicated lines is higher among LLM-specific instructions, which may indicate that such instructions are easier to generalize across different projects or that developers tend to reuse them because they represent a relatively new type of information.

Figure~\ref{fig:duplication-language-distribution} shows the distribution of the average number of duplicated lines for each context type by programming language. We found that different programming languages exhibit different copying practices when writing cursor rules. For example, developers using programming languages for desktop and enterprise applications, such as C\#, which require high domain specificity, tend to replicate cursor rules from similar projects, showing the highest average number of duplicate lines in project-related code (M=61.14, SD=161.77). In contrast, developers using languages for frontend applications, such as JavaScript, often copy pieces from various sources, resulting in evenly distributed duplication across all context types.

For individual repositories that are not complete duplicates of others, the average line duplication rate is 16.30\%, with the top 10 exceeding 96\%. This suggests that some developers extensively reuse existing cursor rules, adding additional context as needed.

A manual investigation of duplicated lines revealed that developers reuse content from diverse sources. Some rules are directly copied from the documentation or cursor rules of dependencies used in the project for project-related context. For example, \texttt{mfts/papermark} copied \textit{``You MUST use \texttt{@trigger.dev/sdk/v3}''} from \texttt{triggerdotdev/trigger.dev}. Developers also reuse guidelines from repositories with similar setups or architectures; for instance, \texttt{github.com/sadmann7/shadcn-table} copies rules like \textit{``Prefer iteration and modularization over code duplication''} from other repositories in the same ecosystem. Additionally, some cursor rules are taken from community-maintained template repositories, such as \texttt{awesome-cursor-rules}, which provide standardized rule sets for common use cases. A small proportion of cursor rule files appear to be generated by LLMs, as indicated by meta-comments or conversational lines. For example, \textit{``Okay, I can help you summarize this pattern for creating a new Cursor rule, focusing on the TanStack Query integration and the preferred error handling strategy.''}

\subsection{RQ4: Context Co-occurrence}

To test whether developers provide broad, multi-aspect context in cursor rules, we analyzed the co-occurrence of context categories within repositories and the number of unique codes per repository.

\mysec{Findings}
\Cref{fig:context-type-venn} illustrates the co-occurrence of context categories in cursor rules. Overall, developers tend to provide comprehensive context for LLMs: 149 repositories (37.16\%) include all four core content categories (excluding examples), and 99 repositories (24.69\%) include all five categories.

Developers prefer to write cursor rules in a style similar to traditional project or collaboration documentation, 
which typically covers best practices and standards for software development~\cite{prana2019categorizing, fronchetti2023contributing}.
In our case, this includes project, conventions, and guidelines category. 
People usually pay the same amount of attention to the tuple combinations of those three context types, 
which were relatively evenly distributed (Project and Convention: 296, 73.8\%; Project and Guideline: 310, 77.3\%; Guideline and Convention: 311, 77.6\%).
Examples are usually provided by the traditional documentation to illustrate other context types and help people use the other context types more effectively as well~\cite{exampleeffectiveness}.  
In our case, the Example category is usually co-occurs with all other three categories: Guideline, LLM Directive, and Convention (172, 42.9\%). Individually, examples are most frequently provided alongside Guidelines (193, 48.1\%), Conventions (189, 47.1\%), LLM Directives (187, 46.6\%)

Some developers choose to only focus on specific aspects when writing cursor rules to let LLM focus 
on a really specific aspects while generating code. 
32 repositories (7.98\%) provided only one category of context in their cursor rules; 
most of them were for guideline-related context (11, 2.74\%).

At a more granular level, developers typically provide a selective number of codes in their cursor rules to avoid overwhelming LLMs and to keep the model focused on the most critical aspects of the project. \Cref{fig:context-type-correlation} shows the distribution of the number of unique codes per repository. Very few repositories provide more than 19 unique codes, and most repositories (323, 80.55\%) include a moderate number of unique codes (between 3 and 15), with a peak at 9 unique codes. This suggests that developers balance context coverage across the project but are selective in their choices, which is consistent with prior findings on documentation practices~\cite{maalej2013patterns}.

\subsection{RQ5: Cursor Rule Composition Over Time}

To answer this research question, we calculated the time difference between the first commit in the cursor rule folder (using the official default directory, i.e., \texttt{.cursor/}) and the repository creation date. Since many repositories did not follow the official default directory structure, we were only able to obtain this information for 281 repositories. We then grouped the repositories based on this time difference into four buckets, ensuring a balanced number of repositories in each group with meaningful time intervals. This resulted in four groups: within seven months (91 repositories), within two years (82), within five years (100), and more than five years (101).

\begin{figure}[t!]
    \centering
    \includegraphics[width=0.4\textwidth]{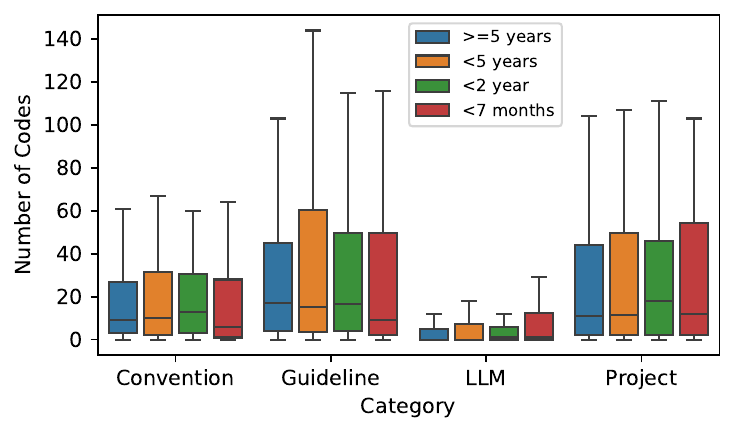}
    \caption{The evolution of the number of codes for each category over time.}
    \label{fig:context-type-before-after-distribution-2}
\end{figure}

\mysec{Findings}
\Cref{fig:context-type-before-after-distribution-2} shows the distribution of the average number of codes for each context type across these time difference groups. Contrary to our initial assumption, developers use similar  strategies when writing project context and convention context in cursor rules, regardless of when the repository was created. 

The most notable trend is observed in the LLM Directives category. As the time difference increases, the amount of LLM-related context shows a slight decrease. Developers of older repositories seem to care less about LLMs and primarily use cursor rules as a form of documentation, focusing more on guidelines and project information. In contrast, newly created repositories tend to be more aware of the boundaries and capabilities of LLMs, and thus provide more LLM-specific instructions.

The amount and type of context included in cursor rules vary systematically with the age of the project relative to cursor rule creation, reflecting changes in developer needs over time.
Groups created within seven months tend to provide the least median amount of all context types, suggesting either they are more concise in their cursor rules overall, or these projects are newer and have less accumulated complexity, requiring less context for LLMs to understand the project effectively.
For groups with a time difference greater than seven months, as the time difference increases, there is a slight decrease in the median amount of project-related and convention-related context provided. This suggests that as projects mature, developers may streamline their cursor rules to focus on the continued maintenance of essential guidelines and LLM instructions, rather than extending current project details.

\mysec{Limitations} 
We only checked the \texttt{.cursor/} folder following official guidance. This may miss monorepos or custom structures, but the bias is consistent and does not affect comparisons.

\section{Discussion and Implications}

\mysec{Enhancing Context Transparency}
Our analysis revealed that a significant portion of cursor rules are duplicates (28.7\% of all lines), indicating that developers often copy and paste content from documentation or community-shared templates that might already be accessible to the coding agent. This practice, along with rules that reference external documents (e.g., ``Consider WordPress coding standards compliance'') without providing links, suggests that developers may be uncertain about the capabilities of AI. This introduces inefficient use of the context budget, which could be better utilized for information not already available to the AI. Enhancing developers' understanding of what context is accessible to the AI (i.e., context transparency) and what information is necessary for the AI, such as providing clearer indications through documentation or UI affordances, would help them provide more targeted and effective context. A real-time feedback mechanism could also be helpful, indicating which rule is currently being used by the coding assistant so that developers can adjust their rules accordingly.

\mysec{Context-Aware Rule Generation} 
We observed that the types of rules that developers think are most effective can vary by programming language or application domain. 
For example, cursor rules for statically typed languages like C\# tend to have more project-specific context, while rules for frontend languages like JavaScript/TypeScript had more examples.
However, developers are currently adopting existing rule templates (28.7\% of all lines) or using LLMs to help generate their cursor rules.
Cursor also officially provides a command to help developers generate initial cursor rules using LLMs. Without proper context awareness, these generated rules may not be optimal for the specific characteristics of each project. 
This indicates significant potential for future tools to generate more \textit{context-aware} rules, tailored to the specific characteristics of each project, rather than relying on generic templates. This could further increase the efficiency of the entire development process. Further research on automatically identifying necessary domain- or language-specific context is needed to inform the design of such tools. Future work can also focus on building better cursor rule generation or context assessment tools to help developers create and evaluate the comprehensiveness of their provided context.

\mysec{Qualifying the Efficacy of Context Rules}
The rules we observed are primarily based on developer intuition; their actual impact on LLM performance remains an open question. For example, it is unclear whether project-related context is more beneficial than allowing LLMs to automatically index project files, as excessive context may degrade LLM performance. Future work is needed to quantify how different types of context—such as the themes in our taxonomy—contribute to task performance and to determine the optimal amount for each context type. Similar to how prompt engineering techniques are empirically tested, the "usefulness" of different context types could be evaluated by comparing end-task results with and without certain context types. This would help prioritize which context is most valuable, especially since authoring and maintaining these rules imposes a cognitive load on developers.

\mysec{Longitudinal Analysis}
Our study provides a cross-sectional snapshot of how developers use cursor rules. However, this is a rapidly evolving practice. As developers become more familiar with LLM capabilities, their use of these rules may change. For example, they may shift from providing detailed project information to focusing more on LLM-specific instructions. Furthermore, as LLMs are upgraded, the relative importance of different context types may shift. A longitudinal study is needed to track the evolution of these rules. One potential approach, common in the MSR community, is to observe whether repositories abandon, expand, or refine their rules as a proxy for the  perceived usefulness and the changing nature of human-AI collaboration in software engineering.

\mysec{Future Education}
Our analysis of creation-time differences shows that newly created repositories often contain less context, suggesting that developers are uncertain about what is necessary. This highlights the need for further education. Our findings across multiple dimensions of developer-provided context can inform educational materials, which should help developers better understand and apply best practices.

\section{Conclusion}

In this paper, we presented a large-scale empirical study of 401 open-source repositories containing cursor rules to understand what types of context developers provide to LLM-powered coding assistants. We identified five main categories of context: Project Information, Convention, Guideline, LLM Directive and Example, with some overlapping information traditionally shared between human developers and others specific to LLMs. We believe our findings, particularly the detailed taxonomy, provide a valuable foundation for the design of future LLM-powered coding assistants, as well as for understanding the knowledge and information that will be shared within teams as LLMs become new collaborators.

\mysec{Data Availability}
Our supplementary material, the dataset and the prompts, is available at \DOIbox{https://doi.org/10.5281/zenodo.18313203}~\cite{jiang_2026_18313203}.

\bibliographystyle{ACM-Reference-Format}
\bibliography{daye}
 
\end{document}